\documentclass[12pt]{article}

\usepackage[cp1251]{inputenc}
\usepackage[T2A]{fontenc}
\usepackage[english]{babel}
\usepackage{amsmath,amsfonts,amssymb}
\usepackage{amsbsy}
\usepackage{bm}

\usepackage{graphicx}
\usepackage{cite}
\usepackage{epstopdf}

\renewcommand{\vec}[1]{\boldsymbol{#1}}

\newcommand {\const} {\mathop {\rm const} \nolimits}

\begin{document}

\begin{center}
\Large\textsc{\textbf{Nonlinear Effects in a Weakly Nonholonomic Systems With a Small Degrees of Freedom}}
\end{center}
{\vskip 0.5cm}
\begin{center}
{\Large\bf Alexander S. Kuleshov${}^{*}$, Nikita M. Vidov${}^{*}$}
{\vskip 0.3cm}
{${}^{*}$ Department of Mechanics and Mathematics, M.~V.~Lomonosov Moscow State University,\\
        Moscow, 119234. \textit{E-mail: kuleshov@mech.math.msu.su, alexander.kuleshov@math.msu.ru}}
\end{center}
{\vskip 1cm}

\begin{abstract}
In 1986 Ya.~V.~Tatarinov presented the foundations of the theory of weakly nonholonomic systems. Mechanical systems with nonholonomic constraints depending on a small parameter are considered. It is assumed that for zero value of this parameter the constraints of such a system become integrable; i.e., in this case, we have a family of holonomic systems depending on several arbitrary integration constants. We will assume that these holonomic systems are completely integrable Hamiltonian systems. When the small parameter is not zero, the behavior of such systems can be considered with the help of normalization methods.

The behavior of such a system can be represented as a combination of the motion of a slightly modified holonomic system with slowly varying previous integration constants (the transgression effect). In this paper we describe the corresponding effects in a several nonholonomic systems with a small degrees of freedom.

\end{abstract}


\subsection{Introduction}

In 1986 Ya.~V.~Tatarinov introduced the concept of weakly nonholonomic systems in his presentation~\cite{Tatarinov1}. The theory of weakly nonholonomic systems was further developed in~\cite{Tatarinov2,Tatarinov3,Tatarinov4}.

Let us consider a mechanical system, whose position is defined by $n+m$ generalized coordinates $x_1,\ldots, x_{n+m}$. Let us assume, that the motion of this system is subjected to nonholonomic constraints:
\begin{equation}\label{1}
\sum\limits_{i=1}^{n+m} a_{si}\left(x_1,\ldots, x_{n+m},\,
\varepsilon\right)\dot{x}_i=0,\quad s=1,\ldots, m,\quad {\rm
rank}\left(a_{si}\right)=n,
\end{equation}
where $\varepsilon$ is a small parameter. Let us assume, that at $\varepsilon=0$ these equations are integrable
\begin{equation*}
\begin{array}{c}
\displaystyle\sum\limits_{s=1}^{m}\displaystyle\sum\limits_{i=1}^{n+m} k_{rs}\left(x_1,\ldots, x_{n+m}\right) a_{si}\left(x_1,\ldots, x_{n+m},\,0\right)\dot{x}_i=\displaystyle\frac{d}{dt}\varphi_r\left(x_1,\ldots, x_{n+m}\right), \\
\det k_{rs}\ne 0, \quad r=1,\ldots, m.
\end{array}
\end{equation*}

Then at $\varepsilon\ne 0$ the such a system will be called weakly nonholonomic system.

Let us introduce the new generalized coordinates $q_1,\ldots, q_{n+m}$ such that $q_i=x_i$, $i=1,\ldots, n$ and
\begin{equation*}
q_{n+\mu}=\varphi_{\mu}\left(x_1,\ldots, x_{n+m}\right),\quad \mu=1,\ldots, m.
\end{equation*}

Then the nonholonomic constraints~\eqref{1}, written in the new variables $q_1,\ldots, q_{n+m}$, have the form:
\begin{equation}\label{2}
\begin{array}{l}
\dot{q}_{n+\mu}=0,\quad \mu=1, \ldots, m,\quad \mbox{at}\quad \varepsilon=0; \\
\dot{q}_{n+\mu}=\varepsilon\displaystyle\sum\limits_{\lambda=1}^n
c_{s\lambda}\left(q_1,\ldots, q_{n+m},\,\varepsilon\right)\dot{q}_{\lambda},\quad
\mu=1, \ldots, m,\quad \mbox{at}\quad \varepsilon\ne 0.
\end{array}
\end{equation}

Let us assume that the Lagrange function is given:
\begin{equation*}
L=L\left(\dot{q}_1,\ldots, \dot{q}_{n+m},\, q_1,\ldots, q_{n+m},\, \varepsilon\right).
\end{equation*}

In this case, at $\varepsilon=0$, we obtain a family of Hamiltonian systems with parameters $R_\mu\equiv q_{n+\mu}$ emerging after integration of constraints~\eqref{2}. The Hamilton function
\begin{equation*}
H_0\left(p_1,\ldots, p_n,\, q_1,\ldots, q_n,\, R_1,\ldots, R_m\right)
\end{equation*}
can be obtained in a standard way from the Lagrange function
\begin{equation*}
\begin{array}{l}
L_0\left(\dot{q}_1,\ldots, \dot{q}_n,\, q_1,\ldots, q_n,\,
R_1,\ldots, R_m\right)=\\ =L\left(\dot{q}_1,\ldots,
\dot{q}_n,\, 0,\ldots, 0,\, q_1,\ldots, q_n,\, R_1,\ldots,
R_m,\, 0\right).
\end{array}
\end{equation*}

At $\varepsilon\ne 0$, the values $R_1,\ldots, R_m$ can start evolving. In Tatarinov’s studies~\cite{Tatarinov1,Tatarinov2,Tatarinov3,Tatarinov4}, this effect was called transgression. The behavior of such systems for $\varepsilon\ne 0$ can be considered with the help of normalization methods~\cite{BrunoBook,Bruno1,Bruno2,Edneral,KuleshovVidov}. In this paper we describe the corresponding effect in the several nonholonomic systems with a small degrees of freedom.

\subsection{Motion of a particle subject to the nonholonomic constraint}

Let a particle of the mass $m$ moves in space. To describe the motion of the particle we introduce the fixed inertial frame $Oxyz$. The position of the particle is completely determined by its coordinates $x$, $y$ and $z$. Let us assume that the particle moves under the action of the potential force with the potential
\begin{equation}\label{3}
V=\frac{c}{2}\left(x^2+y^2\right).
\end{equation}

Moreover, the motion of the particle are subjected to the nonholonomic constraint of the form
\begin{equation}\label{4}
\dot{z}=\varepsilon\left(\dot{x}y-\dot{y}x\right).
\end{equation}

When $\varepsilon=0$, the nonholonomic constraint~\eqref{4} becomes easily integrable and we have
\begin{equation*}
z=C_0=\const.
\end{equation*}

In this case we have the holonomic system with the kinetic energy
\begin{equation*}
T=\frac{m}{2}\left({\dot x}^2+{\dot y}^2\right)
\end{equation*}
and the potential $V$, which is defined by~\eqref{3}. Equations of motion of this holonomic system can be written in the form of Lagrange equations
\begin{equation}\label{5}
\ddot{x}+\omega^2 x=0,\quad \ddot{y}+\omega^2 y=0,\quad \omega^2=\frac{c}{m}.
\end{equation}

General solution of the system~\eqref{5} can be written as follows:
\begin{equation}\label{6}
x=C_1\cos\omega t+C_2\sin\omega t, \quad y=C_3\cos\omega t+C_4\sin\omega t,
\end{equation}
where $C_1$, $C_2$, $C_3$, $C_4$ are arbitrary constants. Thus, in this case the motion of the particle is oscillatory in nature and it occurs in a plane parallel to the $Oxy$-plane and passing through the point $\left(0,\, 0,\, C_0\right)$.

Now let us consider the case $\varepsilon\ne 0$. In this case we will derive equations of motion of the particle by the Gibbs -- Appell method. The energy of accelerations of the particle has the form
\begin{equation}\label{7}
S=\frac{m}{2}\left({\ddot x}^2+{\ddot y}^2+{\ddot z}^2\right).
\end{equation}

From the nonholonomic constraint, having differentiated it once with respect to time, we find
\begin{equation}\label{8}
\ddot{z}=\varepsilon\left(\ddot{x}y-\ddot{y}x\right).
\end{equation}

After substitution the expression~\eqref{8} for $\ddot{z}$ to the energy of accelerations~\eqref{7}, we obtain the final expression for the energy of accelerations
\begin{equation*}
S=\frac{m}{2}\left({\ddot x}^2+{\ddot y}^2+\varepsilon^2y^2{\ddot x}^2+\varepsilon^2x^2{\ddot y}^2-2\varepsilon^2xy\ddot{x}\ddot{y}\right).
\end{equation*}

Equations of motion of the particle, written in the Gibbs -- Appell form, has the form
\begin{equation*}
\frac{\partial S}{\partial\ddot{x}}=-\frac{\partial V}{\partial x},\quad \frac{\partial S}{\partial\ddot{y}}=-\frac{\partial V}{\partial y}
\end{equation*}
or, if we write them explicitly
\begin{equation}\label{9}
\begin{array}{l}
m\left(1+\varepsilon^2y^2\right)\ddot{x}-m\varepsilon^2xy\ddot{y}+cx=0,\\ \\
m\left(1+\varepsilon^2x^2\right)\ddot{y}-m\varepsilon^2xy\ddot{x}+cy=0.
\end{array}
\end{equation}

If we solve the system~\eqref{9} with respect to $\ddot{x}$ and $\ddot{y}$ we obtain from this system
\begin{equation}\label{10}
\ddot{x}=-\omega^2 x,\quad \ddot{y}=-\omega^2 y,\quad \omega^2=\frac{c}{m}.
\end{equation}

General solution of the system~\eqref{10} has the form~\eqref{6}. Substituting expressions~\eqref{6} for $x=x\left(t\right)$ and $y=y\left(t\right)$ to the nonholonomic constraint~\eqref{4}, we reduce it to the form:
\begin{equation*}
\dot{z}=\varepsilon\omega\left(C_2C_3-C_1C_4\right),
\end{equation*}
and therefore
\begin{equation*}
z=\varepsilon\omega\left(C_2C_3-C_1C_4\right)t+C_0.
\end{equation*}

Thus, in the case, when $\varepsilon\ne 0$, the oscillatory motion of the particle in the plane, parallel to the $Oxy$-plane is also accompanied by the motion of this plane along the $Oz$ axis with a constant velocity $v$ of the first order in $\varepsilon$:
\begin{equation*}
v=\varepsilon\omega\left(C_2C_3-C_1C_4\right).
\end{equation*}

Over a time $t$ of the order $t\sim\frac{1}{\varepsilon}$ the system is displaced along the $Oz$ axis from the initial position $z=C_0$ by a finite distance.

\subsection{Almost Holonomic Pendulum}

Let a massless lamina moves in the vertical plane $Oxy$ under the action of gravity. The lamina carries two blades arranged in a {\tt T} shape. The transversal blade slowly moves on a lamina along itself. At the longitudinal blade’s line, a particle $M$ is fixed to the lamina. Let us introduce the fixed coordinate system $Oxyz$ and a moving coordinate system $M\xi\eta\zeta$ rigidly connected with the lamina. Let the origin of the moving coordinate system be located at the point $M$ (where the particle is located); the $M\xi$-axis is directed along the direction of the transversal blade $A$ and the $M\eta$-axis is directed along the direction of the longitudinal blade $B$. The $M\zeta$-axis has the same direction as the $Oz$-axis. Then radius-vectors of points $A$ and $B$ in the coordinate system $M\xi\eta\zeta$ have the form
\begin{equation*}
{\vec r}_A=\xi{\vec e}_{\xi}+d{\vec e}_{\eta},\quad d={\rm const},\quad
{\vec r}_B=r{\vec e}_{\eta},\quad r={\rm const}.
\end{equation*}

Let us assume that the variable $\xi$ varies according to the law $\xi=\xi_0+\varepsilon v t$, where $\varepsilon$ is a small parameter. Let us denote by $C$ the instant center of rotation of the lamina. In the $M\xi\eta\zeta$ coordinate system the radius-vector of the point $C$ has the form:
\begin{equation*}
{\vec r}_C=\xi{\vec e}_{\xi}+r{\vec e}_{\eta}=\left(\xi_0+\varepsilon v t\right){\vec e}_{\xi}+r{\vec e}_{\eta}.
\end{equation*}

\begin{figure}[htb]
\centering
\includegraphics[width=7.5cm]{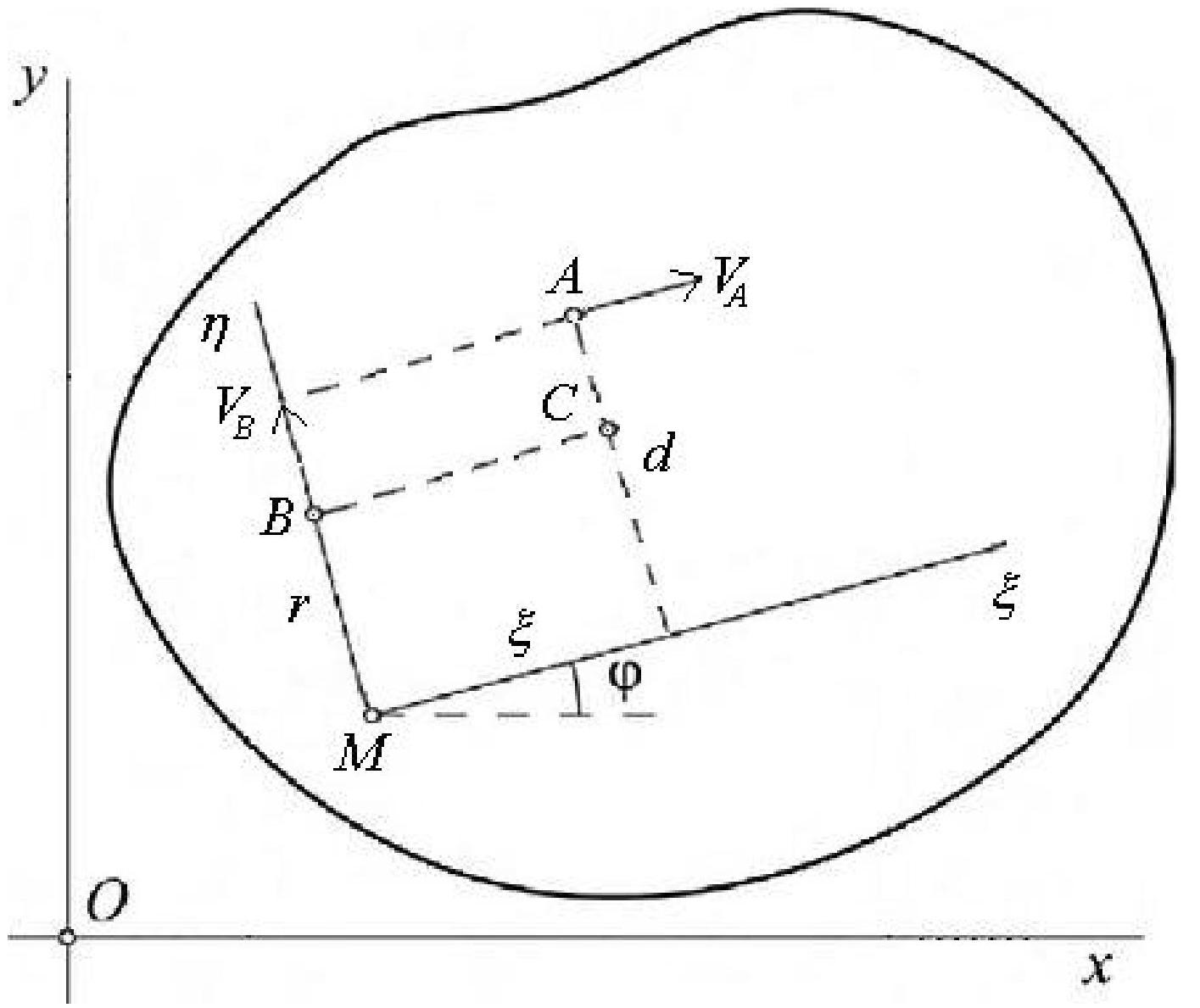}
\caption{Almost Holonomic Pendulum.}
\end{figure}

To describe the position of the system we introduce the coordinates $x$ and $y$ of the point $C$ with respect to the fixed coordinate system $Oxyz$ and the rotation angle $\varphi$ of the lamina (the angle between the $Ox$-axis and the $M\xi$-axis). Then the equations of nonholonomic constraints imposed on the system will have the form
\begin{equation}\label{11}
\dot{x}=\dot{\xi}\cos\varphi=\varepsilon v\cos\varphi,\quad \dot{y}=\dot{\xi}\sin\varphi=\varepsilon v\sin\varphi.
\end{equation}

Equations~\eqref{11} can be rewritten as follows:
\begin{equation}\label{12}
x'=\frac{dx}{d\xi}=\cos\varphi,\quad y'=\frac{dy}{d\xi}=\sin\varphi
\end{equation}
(here, the variable $\xi$ plays the role of slow time). Since the lamina moves under the action of gravity, the point $M$ is subjected to the action of the force ${\vec F}=-Mg{\vec e}_y$.

The dynamic equations of motion of this system can be written in the form of the Gibbs -- Appell equations. With this purpose, let us find first the energy of accelerations of this system. It is defined by the formula
\begin{equation*}
S=\frac{M}{2}{\vec a}_M^2,
\end{equation*}
where ${\vec a}_M$ is the acceleration of the point $M$. In the explicit form, this expression can be written as follows:
\begin{equation*}
S=\frac{M}{2}\left(\left(\xi^2+r^2\right)\left({\ddot{\varphi}}^2+{\dot{\varphi}}^4\right)+{\dot{\xi}}^2{\dot{\varphi}}^2+
2\xi\dot{\xi}\dot{\varphi}\ddot{\varphi}-2r\dot{\xi}{\dot{\varphi}}^3\right).
\end{equation*}

The potential energy of the system has the form
\begin{equation*}
V=Mg\left(y-\xi\sin\varphi-r\cos\varphi\right).
\end{equation*}

The equation describing the change in the angle $\varphi$ can be written as follows:
\begin{equation*}
\frac{\partial S}{\partial\ddot{\varphi}}=-\frac{\partial V}{\partial\varphi}
\end{equation*}
or, in explicit form,
\begin{equation}\label{13}
\left(\xi^2+r^2\right)\ddot{\varphi}+\xi\dot{\xi}\dot{\varphi}+gr\sin\varphi-g\xi\cos\varphi=0.
\end{equation}

Thus, the motion of the almost holonomic pendulum is described by the equations~\eqref{11},~\eqref{13}. Let us make a change of variable in equation~\eqref{13} by using the formula
\begin{equation}\label{14}
\varphi=\psi+\arctan\frac{\xi}{r}.
\end{equation}

After the change of independent variable~\eqref{14} equations of motion of the system take the form:
\begin{equation}\label{15}
\begin{array}{l}
\left(\xi^2+r^2\right)\ddot{\psi}+\xi\dot{\xi}\dot{\psi}+g\sqrt{\xi^2+r^2}\sin\psi-\displaystyle\frac{r\xi\dot{\xi}^2}{\xi^2+r^2}=0, \\ \\
\dot{x}=\displaystyle\frac{r\cos\psi-\xi\sin\psi}{\sqrt{\xi^2+r^2}}\dot{\xi}, \quad
\dot{y}=\displaystyle\frac{r\sin\psi+\xi\cos\psi}{\sqrt{\xi^2+r^2}}\dot{\xi}, \\ \\
\dot{\xi}=\varepsilon v.
\end{array}
\end{equation}

From the first equation of the system~\eqref{15} we obtain that the unperturbed system (corresponding to the value of parameter
$\varepsilon=0$) can be interpreted as a simple pendulum with a length
\begin{equation*}
l=\sqrt{\xi_0^2+r^2}.
\end{equation*}
with the suspension point located at the instant rotation center $C$. Obviously, the unperturbed system has the equilibrium $\psi=0$. In the small vicinity of this equilibrium we will consider the variables $\dot{\xi}$, $\psi$ and $\dot{\psi}$ to be small. Let us expand the right hand sides of equations~\eqref{15} into a power series in $\psi$, $\dot{\psi}$ and $\dot{\xi}$. As a result we obtain:
\begin{equation}\label{16}
\begin{array}{l}
\left(\xi^2+r^2\right)\ddot{\psi}+\xi\dot{\xi}\dot{\psi}+g\sqrt{\xi^2+r^2}\psi-\displaystyle\frac{r\xi\dot{\xi}^2}{\xi^2+r^2}- \displaystyle\frac{g\sqrt{\xi^2+r^2}}{6}\psi^3+ O\left(\varepsilon^5\right)=0, \\ \\
\dot{x}=\displaystyle\frac{r}{\sqrt{\xi^2+r^2}}\dot{\xi}-\frac{\xi}{\sqrt{\xi^2+r^2}}\psi\dot{\xi}-\frac{r}{2\sqrt{\xi^2+r^2}}\psi^2
\dot{\xi}+O\left(\varepsilon^4\right), \\ \\
\dot{y}=\displaystyle\frac{\xi}{\sqrt{\xi^2+r^2}}\dot{\xi}+\frac{r}{\sqrt{\xi^2+r^2}}\psi\dot{\xi}-\frac{\xi}{2\sqrt{\xi^2+r^2}} \psi^2\dot{\xi}+O\left(\varepsilon^4\right),\\ \\
\dot{\xi}=\varepsilon v.
\end{array}
\end{equation}

The first approximation system has the form:
\begin{equation}\label{17}
\ddot{\psi}+\displaystyle\frac{g}{\sqrt{\xi^2+r^2}}\psi=0, \quad
\dot{x}=\displaystyle\frac{r}{\sqrt{\xi^2+r^2}}\dot{\xi},\quad
\dot{y}=\displaystyle\frac{\xi}{\sqrt{\xi^2+r^2}}\dot{\xi},\quad
\dot{\xi}=\varepsilon v.
\end{equation}

We can write the second and the third equation of the system~\eqref{17} as follows:
\begin{equation}\label{18}
\frac{dx}{d\xi}=\displaystyle\frac{r}{\sqrt{\xi^2+r^2}},\quad \frac{dy}{d\xi}=\displaystyle\frac{\xi}{\sqrt{\xi^2+r^2}}.
\end{equation}

Integration of the system~\eqref{18} gives us the following expressions for $x=x\left(\xi\right)$ and $y=y\left(\xi\right)$:
\begin{equation*}
x=r\ln\left(\displaystyle\frac{\xi+\sqrt{\xi^2+r^2}}{r}\right)+C_x, \quad y=\sqrt{\xi^2+r^2}+C_y.
\end{equation*}

From these equations we can find the dependence $y=y\left(x\right)$:
\begin{equation*}
y-C_y=r\cosh\left(\frac{x-C_x}{r}\right).
\end{equation*}

Thus, in the first approximation the motion of an almost holonomic pendulum can be represented as oscillations about a slowly rotating direction of the fixed blade; the instant rotation center $C$ moves along a catenary (transgression effect).

Instead the variables $x$ and $y$ we introduce the new variables $X$ and $Y$ by the formulas
\begin{equation*}
X=x-r\ln\left(\displaystyle\frac{\xi+\sqrt{\xi^2+r^2}}{r}\right),\quad Y=y-\sqrt{\xi^2+r^2}.
\end{equation*}

In the new variables equations of motion of an almost holonomic pendulum~\eqref{16} take the form:
\begin{equation}\label{19}
\begin{array}{l}
\left(\xi^2+r^2\right)\ddot{\psi}+\xi\dot{\xi}\dot{\psi}+g\sqrt{\xi^2+r^2}\psi-\displaystyle\frac{r\xi\dot{\xi}^2}{\xi^2+r^2}- \displaystyle\frac{g\sqrt{\xi^2+r^2}}{6}\psi^3+O\left(\varepsilon^5\right)=0,\\ \\
\dot{X}=-\displaystyle\frac{\xi}{\sqrt{\xi^2+r^2}}\psi\dot{\xi}-\displaystyle\frac{r}{2\sqrt{\xi^2+r^2}}\psi^2\dot{\xi}+
O\left(\varepsilon^4\right),\\ \\
\dot{Y}=\displaystyle\frac{r}{\sqrt{\xi^2+r^2}}\psi\dot{\xi}-\displaystyle\frac{\xi}{2\sqrt{\xi^2+r^2}}\psi^2\dot{\xi}+ O\left(\varepsilon^4\right), \\ \\
\dot{\xi}=\varepsilon v.
\end{array}
\end{equation}

Let us reduce the system~\eqref{19} to the normal form. We will use the method, describing in~\cite{KuleshovVidov}. The normalized 3rd order system has the form:
\begin{equation}\label{20}
\begin{array}{l}
\dot{z}_1=\displaystyle\frac{i\sqrt{g}}{\left(\xi^2+r^2\right)^{\frac{1}{4}}}z_1-\frac{\xi}{4\left(\xi^2+r^2\right)}z_1\dot{\xi}- \displaystyle\frac{i\xi^2}{32\sqrt{g}\left(\xi^2+r^2\right)^{\frac{7}{4}}}z_1\dot{\xi}^2-\frac{i\sqrt{g}}{16\left(\xi^2+r^2
\right)^{\frac{1}{4}}}z_1^2z_2,\\ \\
\dot{z}_2=-\displaystyle\frac{i\sqrt{g}}{\left(\xi^2+r^2\right)^{\frac{1}{4}}}z_2-\displaystyle\frac{\xi}{4\left(\xi^2+r^2\right)}
z_2\dot{\xi}+\displaystyle\frac{i\xi^2}{32\sqrt{g}\left(\xi^2+r^2\right)^{\frac{7}{4}}}z_2\dot{\xi}^2+\displaystyle\frac{i\sqrt{g}}
{16\left(\xi^2+r^2\right)^{\frac{1}{4}}}z_1 z_2^2,\\ \\
\dot{z}_X=-\displaystyle\frac{r\xi^2}{g\left(\xi^2+r^2\right)^2}\dot{\xi}^3-\displaystyle\frac{r}{4\sqrt{\xi^2+r^2}}z_1z_2
\dot{\xi},\quad
\dot{z}_Y=\displaystyle\frac{r^2\xi}{g\left(\xi^2+r^2\right)^2}\dot{\xi}^3-\displaystyle\frac{\xi}{4\sqrt{\xi^2+r^2}}z_1z_2
\dot{\xi}, \\ \\
\dot{\xi}=\varepsilon v,
\end{array}
\end{equation}
where we introduce the following new variables
\begin{equation*}
z_1=\psi-\displaystyle\frac{i\dot{\psi}}{\sqrt{\frac{g}{\sqrt{\xi^2+r^2}}}}+O\left(\varepsilon^2\right)=\bar{z}_2, \quad
z_X=X+O\left(\varepsilon^2\right),\quad z_Y=Y+O\left(\varepsilon^2\right).
\end{equation*}

Now instead of the variables $z_1$, $z_2$ we introduce the new variables $\Delta$ and $\theta$ such that
\begin{equation*}
z_1=\bar{z}_2=\Delta\exp\left(i\theta\right),\quad z_2=\Delta\exp\left(-i\theta\right).
\end{equation*}

In this case
\begin{equation*}
\Delta=\sqrt{z_1z_2},\quad \Delta^2=z_1z_2.
\end{equation*}

Let us derive the differential equation for the new variable $\Delta$. We have
\begin{equation*}
\displaystyle\frac{d\left(\Delta^2\right)}{dt}=2\Delta\dot{\Delta}=\displaystyle\frac{d\left(z_1 z_2\right)}{dt}=\dot{z}_1 z_2 + z_1 \dot{z}_2=-\displaystyle\frac{\xi}{2\left(\xi^2+r^2\right)}z_1 z_2\dot{\xi}=-\displaystyle\frac{\xi} {2\left(\xi^2+r^2\right)} \Delta^2\dot{\xi}.
\end{equation*}

Therefore
\begin{equation}\label{21}
\dot{\Delta}=-\displaystyle\frac{\xi}{4 (\xi^2 + r^2)} \Delta \dot{\xi}.
\end{equation}

Integration of the differential equation~\eqref{21} gives the following expression for the variable $\Delta=\Delta\left(\xi\right)$:
\begin{equation*}
\Delta=C_{\Delta}\left(\xi^2+r^2\right)^{-\frac{1}{8}}.
\end{equation*}

The variable $\theta$ can be found from the following differential equation
\begin{equation*}
\begin{array}{l}
\dot{\theta} = \displaystyle\frac{d}{dt}\left(\frac{1}{2 i} \ln\left(\frac{z_1}{z_2}\right) \right) = \frac{i}{2}\left( \frac{\dot{z}_2}{z_2} - \frac{\dot{z}_1}{z_1} \right) = \\ \\
=\displaystyle\frac{\sqrt{g}}{\left(\xi^2 + r^2\right)^{\frac{1}{4}}}-\frac{\sqrt{g} z_1 z_2}{16\left(\xi^2 + r^2\right)^{\frac{1}{4}}} - \frac{\xi^2 \dot{\xi}^2}{32 \sqrt{g}\left(\xi^2 + r^2\right)^{\frac{7}{4}}},
\end{array}
\end{equation*}

As a result, we obtain for $\theta=\theta\left(t\right)$:
\begin{equation*}
\begin{array}{l}
\theta-\theta_0=\displaystyle\int\limits_0^t\left(\frac{\sqrt{g}}{\left(\xi^2 + r^2\right)^{\frac{1}{4}}}-\frac{\sqrt{g} C_\Delta^2}{16\left(\xi^2 + r^2\right)^{\frac{1}{2}}} - \frac{\varepsilon^2 \xi^2 v^2}{32 \sqrt{g}\left(\xi^2 + r^2\right)^{\frac{7}{4}}}\right)dt.
\end{array}
\end{equation*}

Having the explicit expressions for $\Delta$ and $\theta$ we can find the explicit expressions for $z_1$ and $z_2$. Using these expressions we find the explicit expression for $\psi$:
\begin{equation}\label{22}
\begin{array}{l}
\psi=\displaystyle\frac{\varepsilon C_{\delta}}{\left(\xi^2+r^2\right)^\frac{1}{8}}\cos\theta+\frac{\varepsilon^2 v \xi \left(8rv+ \sqrt{g}C_{\delta}\sin\theta\left(\xi^2+r^2\right)^\frac{5}{8}\right)}{8g\left(\xi^2+r^2\right)^\frac{3}{2}}+\\ \\
+\displaystyle\frac{\varepsilon^3 C_{\delta}\cos\theta\left(g C_{\delta}^2\left(\xi^2+r^2\right)^\frac{5}{4}\left(5+
4\sin^2\theta\right)+6v^2\left(2r^2-\xi^2\right)\right)}{192g\left(\xi^2+r^2\right)^\frac{13}{8}}.
\end{array}
\end{equation}

Now we come back to the equations for $z_X$ and $z_Y$:
\begin{equation*}
\dot{z}_X=-\displaystyle\frac{r\xi^2}{g\left(\xi^2+r^2\right)^2}\dot{\xi}^3-\displaystyle\frac{rC^2_{\Delta}}{4\left(
\xi^2+r^2\right)^\frac{3}{4}}\dot{\xi},\quad
\dot{z}_Y=\displaystyle\frac{r^2\xi}{g\left(\xi^2+r^2\right)^2}\dot{\xi}^3-\displaystyle\frac{C^2_{\Delta}\xi}{4
\left(\xi^2+r^2\right)^\frac{3}{4}}\dot{\xi}.
\end{equation*}

The solution of these equations has the form:
\begin{equation*}
\begin{array}{l}
z_X=-r\displaystyle\int\limits_{\xi_0}^{\xi_0+\varepsilon v t}\left(\frac{\varepsilon^2 v^2\xi^2}{g\left(\xi^2+r^2\right)^2}+ \displaystyle\frac{C^2_{\Delta}}{4\left(\xi^2+r^2\right)^\frac{3}{4}}\right)d\xi+C_X, \\ \\
z_Y =-\displaystyle\frac{r^2\varepsilon^2 v^2}{2g\left(\xi^2+r^2\right)}-\displaystyle\frac{C^2_{\Delta}}{2}\left(
\xi^2+r^2\right)^\frac{1}{4}+C_Y.
\end{array}
\end{equation*}

\begin{figure}[t]
\begin{minipage}[h]{0.49\linewidth}
    \centering
    \includegraphics[width=\linewidth]{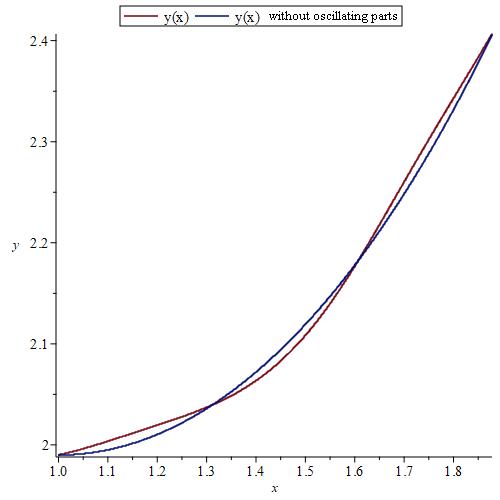}
    \caption{$\varepsilon=0.1$}
\end{minipage}
\begin{minipage}[h]{0.49\linewidth}
    \centering
    \includegraphics[width=\linewidth]{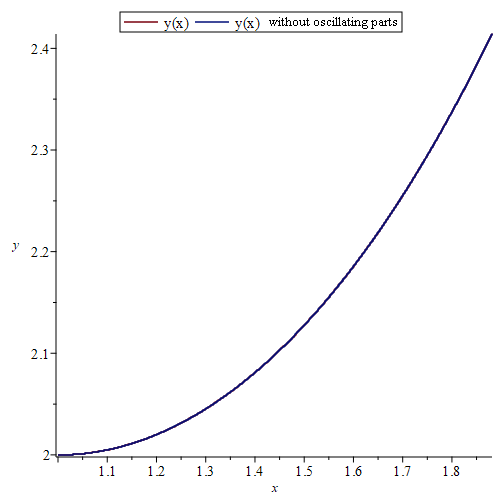}
    \caption{$\varepsilon=0.01$}
\end{minipage}
\end{figure}

Then the explicit expressions for $x$ and $y$ up to the third order terms have the form
\begin{equation}\label{23}
\begin{array}{l}
x=C_X+r\ln\left(\displaystyle\frac{\xi+\sqrt{\xi^2+r^2}}{r}\right)-r\varepsilon^2\displaystyle\int\limits_{\xi_0}^{\xi_0+\varepsilon v t}\left(
\displaystyle\frac{v^2\xi^2}{g\left(\xi^2+r^2\right)^2}+\displaystyle\frac{C^2_{\delta}}{4\left(\xi^2+r^2\right)^\frac{3}{4}} \right)d\xi+f_x\left(\xi,\, \theta\right),\\ \\
y=C_Y+\sqrt{\xi^2+r^2}-\displaystyle\frac{\varepsilon^2}{2}\left(\displaystyle\frac{r^2 v^2}{g\left(\xi^2+r^2\right)}+C^2_{\delta} \left(\xi^2+r^2\right)^\frac{1}{4} \right)+f_y\left(\xi,\,\theta\right).
\end{array}
\end{equation}

Here $C_{\delta}$ is an arbitrary constant such that $C_{\Delta}=\varepsilon C_{\delta}$ and $f_x\left(\xi,\, \theta\right)$, $f_y\left(\xi,\, \theta\right)$ (oscillating parts) have the following form:
\begin{equation*}
\begin{array}{l}
f_x\left(\xi,\,\theta\right)=-\displaystyle\frac{\varepsilon^2 C_{\delta} v}{\sqrt{g}\left(\xi^2+r^2\right)^\frac{3}{8}}
\left(\xi\sin\theta+\varepsilon\cos\theta\displaystyle\frac{v\left(8r^2+\xi^2\right)+2r\sqrt{g}C_{\delta}\sin\theta \left(\xi^2+r^2\right)^\frac{5}{8}}{8\sqrt{g}\left(\xi^2+r^2\right)^{\frac{3}{4}}}\right),\\ \\
f_y\left(\xi,\,\theta\right)=\displaystyle\frac{\varepsilon^2 C_{\delta}v}{\sqrt{g}\left(\xi^2+r^2\right)^\frac{3}{8}}
\left(r\sin\theta-\varepsilon\xi\cos\theta\displaystyle\frac{7rv+2\sqrt{g}C_{\delta}\sin\theta\left(\xi^2+r^2\right)^\frac{5}{8}}
{8\sqrt{g}\left(\xi^2+r^2\right)^{\frac{3}{4}}}\right).
\end{array}
\end{equation*}

Fig.~2--3 show the plot of the function $y=y\left(x\right)$, calculated with oscillating parts $f_x\left(\xi,\, \theta\right)$, $f_y\left(\xi,\, \theta\right)$ (red plot) and without oscillating parts (blue plot) for $\varepsilon=0.1$ (Fig.~2) and for $\varepsilon=0.01$ (Fig.~3). The following values of parameters and initial conditions have been chosen:
\begin{equation}\label{24}
g=1,\quad r=1,\quad v=1,\quad \xi_0=0,\quad C_{\delta}=1,\quad \theta_0=0,\quad C_X=1,\quad C_Y=1.
\end{equation}

We also assume that the time $t$ is changed in the interval $t\in\left[0,\frac{1}{\varepsilon}\right]$.
\begin{figure}[h]
\begin{minipage}[h]{0.49\linewidth}
    \centering
    \includegraphics[width=\linewidth]{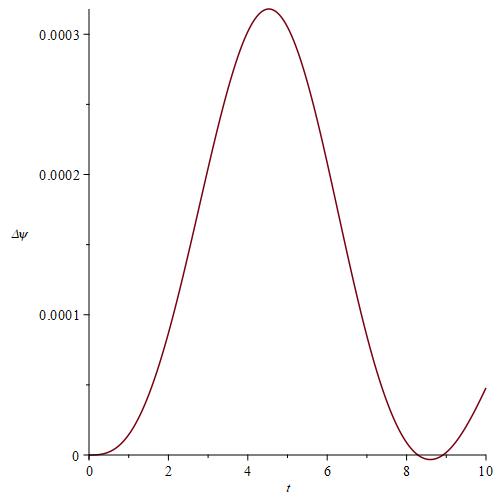}
    \caption{$\varepsilon=0.1$}
\end{minipage}
\begin{minipage}[h]{0.49\linewidth}
    \centering
    \includegraphics[width=\linewidth]{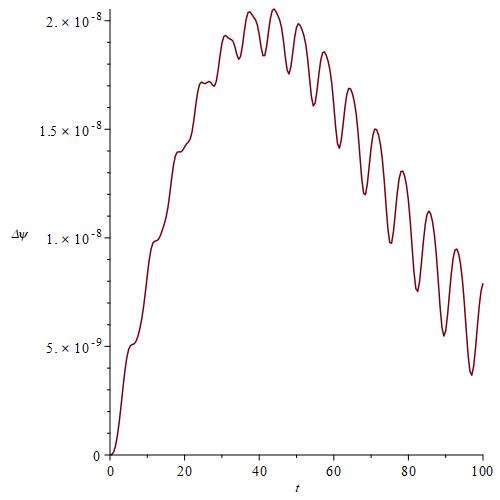}
    \caption{$\varepsilon=0.01$}
\end{minipage}
\end{figure}
\newpage

\begin{figure}[h]
\begin{minipage}[h]{0.48\linewidth}
    \centering
    \includegraphics[width=\linewidth]{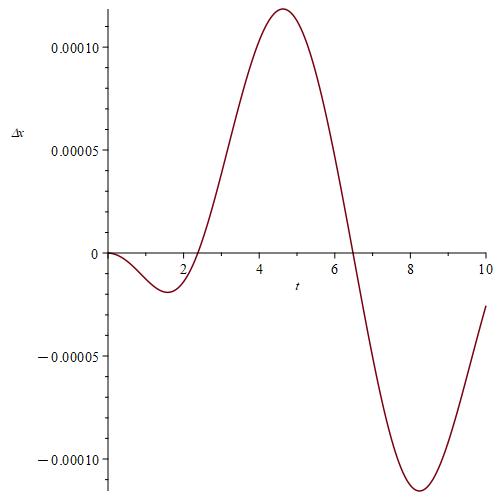}
    \caption{$\varepsilon=0.1$}
\end{minipage}
\begin{minipage}[h]{0.48\linewidth}
    \centering
    \includegraphics[width=\linewidth]{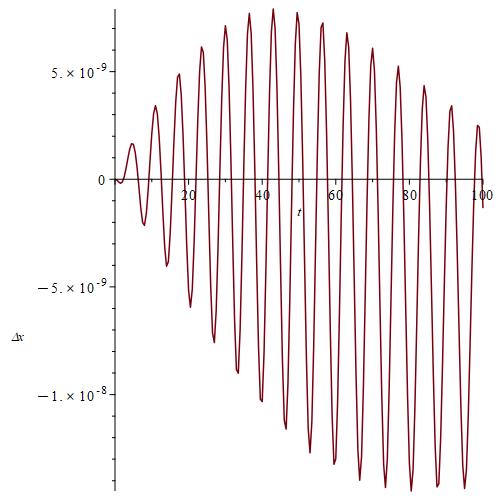}
    \caption{$\varepsilon=0.01$}
\end{minipage}
\begin{minipage}[h]{0.48\linewidth}
    \centering
    \includegraphics[width=\linewidth]{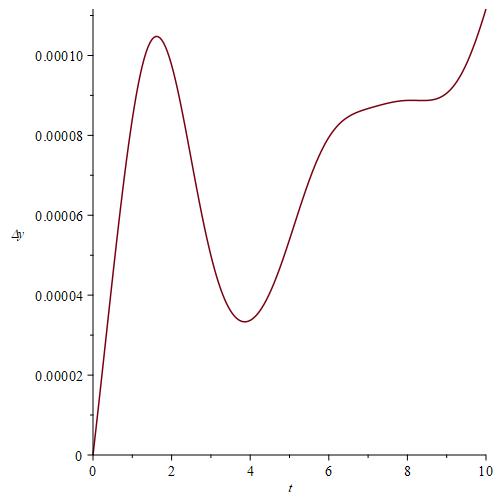}
    \caption{$\varepsilon=0.1$}
\end{minipage}
\begin{minipage}[h]{0.48\linewidth}
    \centering
    \includegraphics[width=\linewidth]{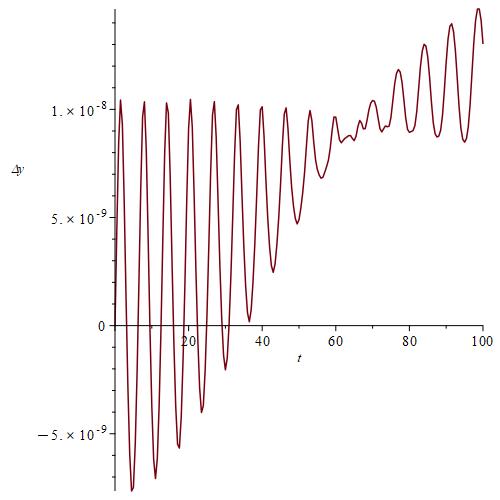}
    \caption{$\varepsilon=0.01$}
\end{minipage}
\end{figure}

Now let us perform the numerical integration of the system~\eqref{15} with the values of parameters and initial conditions~\eqref{24}. Fig.~4--9 show the plot of $\Delta\psi=\psi_{\text{numeric}}-\psi_{\eqref{22}}$, $\Delta x=x_{\text{numeric}}-x_{\eqref{23}}$ and $\Delta y=y_{\text{numeric}}-y_{\eqref{23}}$ for $\varepsilon=0.1$ and $\varepsilon=0.01$, where $t\in\left[0,\frac{1}{\varepsilon}\right]$. Here we denote $\psi_{\eqref{22}}$, $x_{\eqref{23}}$ and $y_{\eqref{23}}$ the functions $\psi$, $x$ and $y$, calculated according to approximate formulas~\eqref{22},~\eqref{23}.

It is evident from the given plots that when $\varepsilon$ decreases by a factor of $10$, the values $\Delta\psi$, $\Delta x$ and $\Delta y$ decreases by a factor of $10^4$. When $t\in\left[0,\frac{1}{\varepsilon}\right]$ the values $\Delta\psi$, $\Delta x$ and $\Delta y$ are fourth order in $\varepsilon$. This allows us to assume that the solution, calculated up to terms of the 3rd order in $\varepsilon$, has been found correctly.

\subsection*{Conclusions.}

In this short paper we considered two examples of nonholonomic systems with a small degrees of freedom in which the transgression effect takes place. This effect can also be found in many other nonholonomic systems. We plan to continue our investigation in this field.

\end{document}